\documentclass[preprint,article,submit,pdftex,moreauthors]{Definitions/mdpi}

\usepackage{soul}

\firstpage{1}
\pubvolume{1}
\issuenum{1}
\articlenumber{0}
\pubyear{2025}
\copyrightyear{2025}
\datereceived{ }
\daterevised{ }
\dateaccepted{ }
\datepublished{ }
\hreflink{https://doi.org/}

\Title{The Lives and Deaths of Red Supergiants}

\TitleCitation{Red Supergiants}

\Author{Emma R. Beasor $^{1}$\orcidA{}}

\AuthorNames{Emma Beasor}

\isAPAStyle{%
       \AuthorCitation{Beasor, E.~R.}
         }{%
        \isChicagoStyle{%
        \AuthorCitation{Beasor, Emma R.}
        }{
        \AuthorCitation{Beasor, E.R.}
        }
}

\address{%
$^{1}$ \quad Liverpool John Moores University; E.R.Beasor@ljmu.ac.uk\\}

\abstract{Red supergiants (RSGs) represent a brief yet pivotal phase in the evolution of massive stars, encompassing the final stages of nuclear burning for many core-collapse supernova progenitors. Having expanded to 100 times larger than their main-sequence sizes, they occupy the cool, luminous extreme of the Hertzsprung--Russell diagram. Characterised by extended convective envelopes, complex and dynamic atmospheres, and, sometimes, substantial mass loss, RSGs play a critical role in shaping both their own evolutionary outcomes and the circumstellar environments into which they ultimately explode. Despite decades of study, fundamental aspects of RSG physics remain poorly constrained, including the mechanisms driving mass loss and the connection between observed RSG populations and the properties of Type II supernovae. Recent observational advances have transformed the field, revealing episodic mass-loss events, asymmetric circumstellar structures, and signs of rapid evolutionary change in individual objects. Alongside these advances, studies of Type II supernovae and their progenitors have raised important new questions about the final stages of massive star evolution, including the possibility of rapid changes in the years immediately preceding core collapse. Motivated by these developments, this review aims to describe our current understanding of how RSGs live and die, from their observed properties through to their role as supernova progenitors, and where that understanding remains incomplete. }


\newcommand{\Msun}{\ensuremath{M_{\odot}}}
\newcommand{\Rsun}{\ensuremath{R_{\odot}}}

\newcommand{\Teff}{\ensuremath{T_{\rm eff}}}        
\newcommand{\logL}{\ensuremath{\log(L/L_{\odot})}} 
\newcommand{\Mdot}{\ensuremath{\dot{M}}}

\newcommand{\kms}{km\,s$^{-1}$}
\newcommand{\Myrunit}{\ensuremath{M_{\odot}\,\mathrm{yr}^{-1}}}

\usepackage[normalem]{ulem}

\begin{document}
\nolinenumbers

\section{Introduction}
For the majority of massive stars, the red supergiant (RSG) phase represents the final stage of evolution before core collapse, with more than half ending their lives as Type~II-P or II-L core-collapse supernovae \citep[CCSNe,][]{smith2011rates}. During this phase, stars expand dramatically to radii often exceeding 1000\,\Rsun, reaching luminosities of $\logL \lesssim 5.5$ and effective temperatures of $\Teff \sim 3500$--$4000$\,K. Although the RSG phase is brief ($\lesssim$10\% of the main sequence lifetime), its impact on the surrounding environment can be profound. Powerful stellar winds shape the circumstellar medium and contribute processed material to the local ISM, while the eventual core-collapse explosion enriches the surrounding environment with heavy elements. RSGs are therefore an important link in the chain connecting massive star formation to the chemical enrichment of galaxies.

The RSG phase as a distinct stage of stellar evolution has been recognised since the 1940s, when \citet{Morgan1943} established luminosity-based spectral classification, which for the first time separated M-type supergiants (luminosity class I) from the morphologically similar giants. The theoretical framework placing RSGs as core helium-burning massive stars was developed in the late 1960s, most notably by \citet{Stothers1968}, who constructed evolutionary models for massive stars in the He-burning phase, and by \citet{Stothers1969}, who surveyed RSGs across Galactic and Magellanic Cloud clusters and discussed the constraints their numbers placed on massive star evolution. Evidence that RSGs were active mass-losing objects came with the development of infrared astronomy in the 1960s and 70s: \citet{Woolf1969} identified silicate emission features at $\sim$10\,$\mu$m in cool M-type supergiants including one of the nearest RSGs to the Sun, $\alpha$~Ori (i.e., Betelgeuse), demonstrating the presence of warm circumstellar dust shells and providing the first direct evidence for significant mass loss in this class of star. The connection to core-collapse supernovae was established purely theoretically, at first, through light-curve modelling \citep[e.g.][]{Grassberg1971, Chevalier1976}, which showed that the extended, hydrogen-rich envelopes characteristic of RSGs naturally produce the plateau morphology of Type~II-P supernovae. This picture was confirmed directly by \citet{VanDyk2003} and \citet{Smartt2004}, who independently identified an RSG at the explosion site of SN~2003gd.

In recent decades, advances across observational and theoretical fronts have transformed our understanding of RSGs. Large-scale photometric surveys have enabled the systematic identification of RSG populations across the Local Group, revealing thousands of candidates where previously only dozens were catalogued \citep{Massey2003}, while the serendipitous archival coverage of nearby galaxies by the \textit{Hubble Space Telescope} has made direct pre-explosion progenitor identification increasingly routine \citep{Smartt2009, VanDyk2025}. Simultaneously, dedicated all-sky transient surveys have substantially increased the volume-limited sample of Type~II CCSNe available for statistical study. On the theoretical side, the development of community stellar evolution codes such as MESA \citep{Paxton2011} has enabled exploration of massive star interiors across wide parameter spaces, incorporating rotation, convective mixing, and mass-loss prescriptions in a single framework. Yet these same advances have sharpened, rather than resolved, a number of fundamental tensions. Modern   measurements of RSG mass-loss rates find that older mass-loss rate prescriptions used in stellar models likely overpredict the total mass shed during the RSG phase by up to a factor of 30 \citep[e.g.][]{Beasor2020, Decin2024}, undermining a core assumption in evolutionary calculations and raising the question of what mechanism, if not quiescent winds, is responsible for determining how stars in the 8--30\Msun\ range end their lives. These conflicts between theory and observation, and their consequences for our understanding of supernovae, form the primary motivation for this review.

The remainder of this review is organised as follows. Section~\ref{sec:evolution} sets out the observed properties of RSGs, their evolutionary context, and the influence of binarity. Section~\ref{sec:massloss} reviews mass-loss measurements and the prescriptions derived from them, and argues that the mass shed during the RSG phase is substantially smaller than stellar evolution models assume. Section~\ref{sec:progenitors} turns to RSGs as supernova progenitors and re-examines the initial mass distribution of SN progenitors in light of that conclusion. 


\section{Properties of Red Supergiants}\label{sec:evolution}

\subsection{Evolutionary Context}

Within the single-star evolutionary paradigm, all massive stars between roughly 8 and 30\,\Msun\ are expected to undergo an RSG phase before ending their lives as core-collapse supernovae (CCSNe) \cite[e.g.][]{Meynet2015}. RSGs occupy the upper right region of the Hertzsprung--Russell diagram. Following core hydrogen exhaustion, hydrogen shell burning ignites and the core contracts. Initially both the core and envelope contract, producing the characteristic ``hook'' in evolutionary tracks, before the envelope reverses and expands. Once the envelope becomes fully convective, the star approaches the Hayashi limit --- the coolest effective temperature at which a star in hydrostatic equilibrium can remain in radiative--convective balance. Because this limit depends only weakly on stellar mass, RSGs of widely differing masses are confined to a narrow range in $\Teff$, and the RSG branch appears in the HRD as a near-vertical sequence spanning more than an order of magnitude in luminosity.

Two practical difficulties complicate the identification of RSGs as a population using the HRD only. First, effective temperatures remain contested: values derived from TiO band strengths are systematically cooler than those from SED fitting or interferometry, by several hundred K \citep{Levesque2005, Davies2013, gonzaleztora2021}, because the TiO bands form far out in an extended atmosphere that is poorly represented by a single photospheric temperature. \citet{dewit2024} provide linear relations between the temperatures derived from TiO bands and from the $J$- and $i$-band spectral regions, and find that the discrepancy between TiO-based and other temperatures increases with mass-loss rate, consistent with TiO absorption forming partly in the wind \citep{daviesplez2021}.
Second, the boundary between RSGs and luminous AGB stars is not sharp (see Figs.~6 and 7 from \citep{massey2021}). At the low-luminosity end, the two populations overlap in colour and magnitude. Without spectra and well-determined effective temperatures \citep[since AGB stars extend to much cooler than RSGs, with a maximum \Teff\ of $\sim$3000~K, versus average $\Teff$ of $\sim$4000~K for RSGs, see][]{hofner2018agbs,Davies2013,gonzaleztora2021} low-mass RSGs cannot be separated from high-mass AGBs photometrically. Both issues complicate efforts to determine luminosity functions, mass-loss rates, and SN progenitor masses.

\subsection{Spectral Characteristics}

The spectra of RSGs are dominated by prominent molecular absorption features, most notably titanium monoxide (TiO), carbon monoxide (CO), and other molecules \cite[see][and refs within]{ziurys2025review}. These strong molecular signatures are a direct consequence of the cool photospheric temperatures, which allow molecular species to survive in the distended stellar atmospheres. The presence and strength of these features provide crucial diagnostics for identifying and classifying RSGs. In the optical, the depth of the TiO bands deepens monotonically along the M-type sequence and forms the basis of the spectral classification. In the near-infrared, the CO bandheads near 2.3\,$\mu$m and the Ca~II triplet at 0.85\,$\mu$m are the clearest indicators of the cool, extended atmospheres of RSGs.

\begin{figure*}
    \centering
    \includegraphics[width=0.9\linewidth]{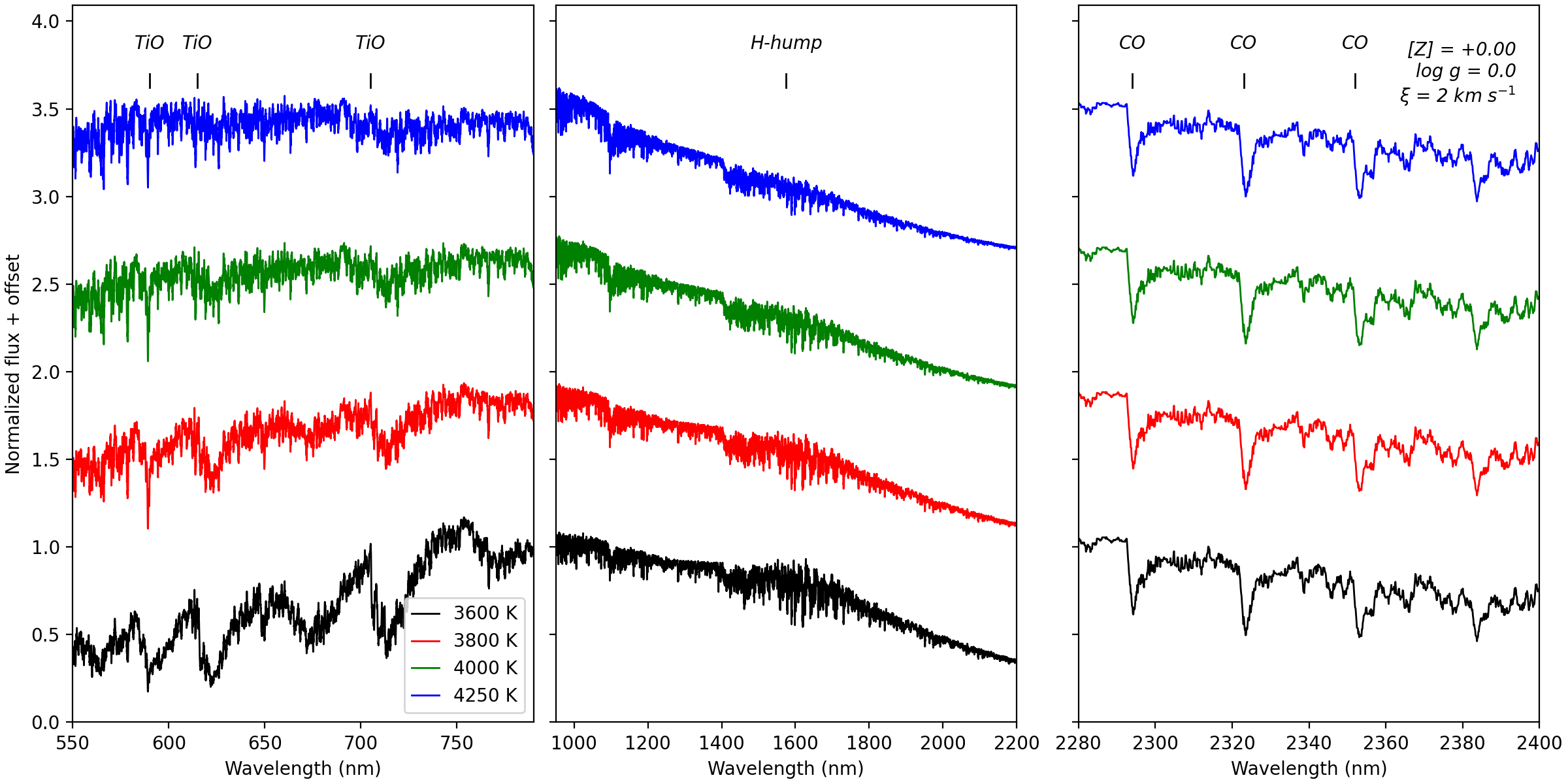}
    \caption{        MARCS spherical model spectra \citep[][]{Gustafsson2008} of red supergiant atmospheres over a
        sequence of effective temperatures, $T_{\rm eff} = 3600$, $3800$,
        $4000$, and $4250\,$K, at fixed surface gravity ($\log g = 0.0$), solar metallicity ($[\mathrm{Z}] = 0.00$), and microturbulent velocity ($\xi = 2\,\mathrm{km\,s^{-1}}$). Spectra are continuum-normalized and vertically offset for clarity, and smoothed with a 10-pixel boxcar filter.
        \textit{Left}: the optical TiO absorption bands (labeled), which
        deepen with decreasing $T_{\rm eff}$ as cooler
        photospheres favor TiO molecule formation.
        \textit{Middle}: the near-infrared continuum shape, including the
        broad H$^{-}$ opacity ``hump'' near $1.5$--$1.6\,\mu\mathrm{m}$
        (labeled).
        \textit{Right}: a zoomed view of the CO first-overtone bandhead
        triplet near $2.29$--$2.35\,\mu\mathrm{m}$ (labeled), which
        similarly strengthens toward lower $T_{\rm eff}$. }
    \label{fig:marcs_teff}
\end{figure*}
In practice, RSGs are usually selected photometrically before any spectroscopy is obtained \citep[e.g.][]{massey2021}, using colour--magnitude cuts in the near-infrared where the effects of interstellar extinction are reduced. The dominant contaminants are foreground Galactic dwarfs, which can be removed using proper motions and parallaxes, and luminous AGB stars, which cannot be separated so cleanly and require either a luminosity cut or a spectroscopic follow-up campaign. 

\subsection{The Maximum Luminosity of Red Supergiants}
\label{sec:hdlimit}

A critical empirical constraint on RSG properties is the maximum luminosity observed among RSGs in the local universe. \citet{Stothers1969} and \citet{sandage1974} both noted a lack of cool supergiants above a certain luminosity, while \citet{Humphreys1979}, using Galactic and LMC samples, placed an upper boundary on the coolest supergiants in the HRD at $\log L \simeq 5.8$; this threshold is now commonly known as the Humphreys--Davidson (H--D) limit. More recent work has revised this value downward. Using homogeneously determined luminosities for RSGs in the LMC and SMC, \citet{Davies2018b} found a maximum luminosity of $\logL = 5.5$ in both galaxies, 0.3\,dex below the original estimate, with the difference driven largely by improved bolometric corrections and the removal of foreground and AGB contaminants from earlier samples. The lower value has since been supported at solar metallicity by \citet{McDonald2022}, with no RSG having been observed to exceed $\logL = 5.5$ despite surveys spanning a factor of $\sim$5 in metallicity.

The physical origin of this limit remains unclear. Stellar evolutionary models predict that stars more massive than $\sim\!30 \, \Msun$ experience such substantial envelope mass loss during the main sequence that they never reach the RSG phase. Under this scenario, the H--D limit might be expected to depend on metallicity, since the strength of radiation-driven stellar winds scales with metallicity: lower metallicity would produce weaker winds, lower mass-loss rates, and potentially allow more massive stars to retain their envelopes and reach the RSG phase. However, observations show the H--D limit to be essentially constant from SMC to solar metallicity, independent of the stellar environment or host galaxy. This invariance suggests that a physical mechanism other than metallicity-dependent winds establishes the upper luminosity boundary for RSGs.

RSG winds are thought, in part, to be driven by radiation pressure acting on dust opacity, and since dust formation is expected to depend on metallicity, mass-loss rates might be expected to scale with $Z$ as well. Surprisingly, however, modern surveys of RSG mass-loss find little to no significant dependence on metallicity \citep{Beasor2020, Antoniadis2024, vanLoon2025}. This may indicate that the mass loss is not primarily dust-driven, for example if the wind is initiated by turbulent pressure in the extended atmosphere \citep{Kee2021}, or that a significant fraction of the mass is lost in episodic events \citep[e.g.][]{Cheng2024, bonanos2024episodic, munoz2026} rather than a steady wind (Section~\ref{sec:massloss}).

\subsection{The Influence of Binarity}
\label{sec:binarity}
 
The theoretical picture of RSG evolution, including the interior physics discussed in this and the next section, is reviewed in detail in a recent review by \citet{Ekstrom2025}. Here, we focus on the observational constraints.
 
Most massive stars are born in binaries. Spectroscopic monitoring of the O star population in 30~Doradus finds that the majority have companions close enough to interact at some point in their lives \citep{Sana2013}. Surveys of RSGs find far fewer. Searches for the ultraviolet excess produced by a hot companion give binary fractions of around 30\% in the LMC, dropping further at SMC metallicity ($\sim$20\%)  \citep{Patrick2020, Neugent2021}. 
 
Part of the difference is due to detection limits. Ultraviolet excess measurements are only sensitive to companions above a certain mass and temperature. Radial velocities are of limited use for RSGs, since a B-type companion on a wide orbit produces a reflex motion of a few \kms{} over years, comparable to the velocity variations caused by convection at the surface of the star \citep{Patrick2020}. The measured binary fractions are therefore lower limits, and some of the apparent drop with metallicity may not be real.
 
The remaining difference is physical. An RSG is two to three orders of magnitude larger than the main-sequence star it evolved from, with the most extended objects reaching $\sim$$10^{3}$\,\Rsun{} \citep[e.g.][]{Wittkowski2012}. A star of this size fills its Roche lobe at separations of a few thousand \Rsun, corresponding to orbital periods of a few years. Systems closer than this cannot survive the expansion intact, and in most cases they stop being detectable as binaries as a result.
 
There are three ways this happens. Where the two components merge on the main sequence, the merger product evolves as if it were a single star of a higher initial mass. The product is a blue straggler, which can later evolve into an RSG that is too luminous for the age of its cluster, and mergers of this kind are expected to contribute to the luminosity spread seen among RSGs in young clusters  \citep{Beasor2019, Britavskiy2019, guzman2025, wang2025rsgclusters}. Where mass transfer is stable, the primary is stripped, and the star seen in the red is the secondary. The primary will already have exploded, leaving either a compact companion or, if the system was disrupted, a single RSG with a peculiar velocity \citep{Renzo2019}. Only the widest systems (periods longer than $\sim$10\,yr, or a few $\times 10^3$\,d) survive the RSG phase as ordinary binaries, and it is essentially these that the surveys are counting. A binary fraction of 30\% is therefore a statement about how many RSGs still have detectable companions, not how many were born with one. In addition, many of the wide companion systems are likely originally triples, and the inner binary has already merged \citep{moe2017binaries}.

This matters for the interpretation of RSG samples. If a significant number of apparently single RSGs are the products of binary interaction, then converting an observed luminosity into an initial mass using single-star tracks will give the wrong answer for some fraction of any sample \citep[see e.g.][]{Beasor2019, guzman2025}.  A star that has accreted material, or is the product of a merger, has a luminosity set by its core and total mass after the interaction rather than at birth. The same applies to ages derived for clusters containing RSGs, and to the progenitor masses discussed in Section~\ref{sec:progenitors}, which are derived in exactly this way. This is a systematic error that better photometry cannot remove, and it is not usually included in the quoted uncertainties.
 
Binary interaction also changes the appearance of the star at explosion. A merger following a common envelope phase remains the most likely progenitor scenario for SN\,1987A \citep{Podsiadlowski1990}. Population synthesis suggests that a significant fraction of hydrogen-rich CCSNe come from stars that spent little or no time as an RSG in the single-star evolution sense \citep{Zapartas2019, Schneider2024}.

\subsection{Post-RSG Evolution and Blue Loops}
\label{sec:postRSG}
 
Not all RSGs end their lives in the red. What happens next depends on the mass lost during the RSG phase, as well as on the initial mass and metallicity. Below approximately 12\,\Msun, some single star models predict that a star arriving on the RSG branch will move back towards higher temperatures (i.e. a "blue loop") for part of core helium burning before returning to the red \citep{Anderson2016}. The excursion begins once the hydrogen burning shell reaches the composition discontinuity left by first dredge-up, and its extent depends on how steep that discontinuity is, with a steep gradient suppressing the loop \citep{Walmswell2015}. Loops become wider with increasing initial mass and then stop abruptly, so the boundary between stars that loop and stars that do not is narrow. If the loop is wide enough to reach the classical instability strip the star will pulsate as a Cepheid. 

For more massive RSGs, blue loops in stellar models depend almost entirely on the assumptions made on mass-loss during the RSG phase itself. Once enough mass has been lost that only $\sim$0.1\,\Msun\ remains of the envelope \citep[][]{farrell2022, Meynet2015}, the star contracts and moves to higher temperatures \citep{Giannone1967}. It will appear first as a yellow super/hypergiant, and if stripping continues, eventually as a Wolf-Rayet star \citep{Meynet2015}. The initial mass at which this happens is not well determined. Depending only on which mass-loss prescription is used, evolution to the blue occurs for stars of around 15\,\Msun{} \citep{Georgy2012}, or does not occur at all below about 30\,\Msun{} \citep{Beasor2021}. A factor of two difference caused by a choice of input rather than by any difference in physics illustrates why the mass-loss rates discussed in Section~\ref{sec:massloss} matter.
 
As of yet, above the classical Cepheid mass regime, no luminous post-RSG star has yet been confirmed as a blue loop object in the strict single-star sense. The first problem is that the models are not robust. Loop extent and duration depend on the treatment of convective overshooting, semiconvection and the depth of first dredge-up, none of which are well constrained. The sensitivity is severe enough that different evolutionary codes predict qualitatively different loop behaviour from the same initial conditions \citep[e.g.][]{Tang2014, Martins2013}.
 
The second problem is that a yellow or blue supergiant found where a loop is predicted could have arrived there in several other ways. Binary stripping is the most important of these. A companion on an intermediate period orbit can remove part of the hydrogen envelope by stable mass transfer, stopping or reversing the evolution to the red before the star becomes an RSG, or interrupting an RSG phase that is already underway \citep{Podsiadlowski1992, Eldridge2008}. Near complete stripping leaves a hot, hydrogen-poor star. Partial stripping leaves a star at intermediate temperature that looks much like a loop star but has a different internal structure and history. In this case, a star may not show any sign of a companion if mass transfer has finished, if the companion is compact, or if the binary was disrupted by an earlier explosion \citep{Renzo2019, Zapartas2019}. Given the binary fractions discussed in Section~\ref{sec:binarity}, this is not an unlikely alternative.
 
The two cases can in principle be separated. A star that has been to the red and back should show CNO-processed material at its surface, but stripping by mass transfer can also expose CN-processed layers, so abundances must be interpreted alongside other constraints. The surface abundances and pulsational properties of variable blue supergiants have been used to argue for a post-RSG interpretation on these grounds \citep{Saio2013, Georgy2012}. In practice this requires high resolution spectroscopy and a pulsation analysis for each star, which is why so few identifications have been made. Some yellow supergiants have been suggested as post-RSG candidates \citep[e.g.][]{gordon2016, koumpia2026}, largely on the basis of signatures of mass loss and circumstellar dust. An infrared excess alone is not diagnostic, however, as it is expected both from mass lost during an RSG phase and from non-conservative mass transfer, in which material lost through the outer Lagrange point forms a dusty circumbinary outflow \citep{pejcha2016, lu2023}. Several dusty yellow hypergiants have indeed turned out to be interacting binaries. HR~5171~A has been resolved interferometrically as a binary \citep{chesneau2014}, and HD~144812 shows double-peaked emission from an accretion disc around a hot companion, indicating ongoing mass transfer \citep{kourniotis2025}. Until abundances, pulsation properties and constraints on binarity are available for a reasonable sample, the relative contribution of post-RSG evolution and binary stripping to the population of intermediate temperature supergiants will remain unknown. This also affects the interpretation of yellow supergiant SN progenitors, discussed in Section~\ref{sec:progenitors}.

Stars that keep their envelopes do neither of these things. They reach core collapse in the red and explode as Type II-P SNe \citep{Beasor2021}, or maybe, collapse directly to a black hole without a bright transient (see discussion in Section \ref{sec:progenitors}).
 
\section{Mass-loss rates}
\label{sec:massloss}

All stars lose mass as they evolve, but for massive stars, the amount and timing of this mass loss can have a profound impact on their eventual fate. In particular, its strength determines the type of supernova (SN) they produce at core collapse. Stars that retain most of their hydrogen envelopes will explode as hydrogen-rich, or ``unstripped'', SNe --- typically Type II-P, and sometimes Type II-L. By contrast, stars that shed substantial portions of their envelopes produce hydrogen-poor, or ``stripped'', SNe, such as Type Ib, IIb and Ic. However, the mass loss responsible for producing stripped SNe may not originate from RSG winds at all, but instead from binary interaction, which complicates efforts to understand the specific role of RSG winds in this process. Since RSGs are defined in part by their extended, hydrogen-rich envelopes, the mass lost during this phase sets both the star's subsequent evolutionary track and the observable characteristics of the resulting supernova. Despite its importance, mass loss remains one of the largest sources of uncertainty in massive star evolution.

The mass-loss rates for RSGs specifically have remained a topic of debate for many decades. Part of the issue is that we do not know the driving mechanism for their winds, and so mass-loss rates cannot be derived from first principles. While for hot stars mass loss is comparatively well understood \citep[despite recent conjecture over clumping and the weak-wind problem, e.g. ][]{bjorklund2023,vink2026} through radiation-driven winds acting on metal lines in the stellar photosphere, RSGs present a fundamentally different challenge. As discussed by \citet{vanLoon2025}, this is in part due to RSGs not being themselves a uniform class, with different wind-driving mechanisms likely operating at each end of the temperature range. At the stellar photosphere itself, there is not yet sufficient dust present for radiation pressure to act on, so this mechanism alone cannot lift material from the photosphere up to the dust condensation radius, which typically lies at several stellar radii where temperatures drop below $\sim$1000--1500\,K. Consequently, another mechanism such as large-scale pulsations \citep{Yoon2010} or Alfv\'en-wave driving \citep{Cranmer2011} is required to levitate material into the region where dust can form and radiation pressure can efficiently accelerate the wind. A recent attempt to model the wind-driving mechanism for RSGs from first principles was presented by \citet{Kee2021}, who proposed that turbulent pressure in the extended atmosphere, combined with radiation pressure on dust, can drive the wind. We note that the derived mass-loss rate was highly sensitive to the turbulent velocity, with a change of only a few km/s resulting in a change in mass-loss of many orders of magnitude \citep[see Fig. 3 within][]{Kee2021}. While this work represents significant initial steps, the physical picture remains incomplete, particularly for the most extreme mass-loss episodes observed in some RSGs. 

Spectro-interferometry provides an independent constraint. VLTI/AMBER observations of several RSGs reveal molecular layers of CO and H$_2$O extending beyond the continuum photosphere, comparable to those seen in Mira variables. Neither hydrostatic model atmospheres, 3D convection simulations nor 1D pulsation models reproduce these extensions \citep{Wittkowski2012, arroyotorres2013, arroyotorres2015}, see also \citet{climent2020}. Pulsation and convection alone therefore appear insufficient to levitate material to the radii where molecules and dust form. Semi-empirical models that add a wind to a MARCS atmosphere can reproduce the observed extensions \citep{daviesplez2021, gonzaleztora2023, gonzaleztora2024}, but do not by themselves identify the driving mechanism.

Since it is not yet possible for stellar models to calculate mass loss from first principles, stellar evolution codes require input in the form of mass-loss rate prescriptions that are derived entirely from observations. These empirical prescriptions provide a relation between mass-loss rate and stellar parameters such as luminosity, effective temperature, metallicity and in some cases, stellar mass. The models then apply these prescriptions at each time step, calculating the instantaneous mass-loss rate based on the star's current properties, and evolving the stellar structure accordingly. The total mass lost over the RSG phase --- and therefore the final hydrogen envelope mass at core collapse --- depends sensitively on the choice of prescription, and there are a vast number of prescriptions to choose from, each calibrated on different observational datasets and employing different functional forms \citep{deJager1988, Nieuwenhuijzen1990, vanLoon2005, Goldman2017, Beasor2020, Kee2021, Antoniadis2024, Decin2024}.

Observations of RSG mass-loss rates span several orders of magnitude, from relatively modest winds of $\sim$$10^{-7}$\,\Myrunit{} to episodic, intense outflows at rates exceeding $10^{-4}$\,\Myrunit\, which if sustained for $10^{4}$\,yr would remove of order 1\,\Msun. Observationally, constraining RSG mass loss requires a multi-wavelength approach, combining spectroscopy, infrared dust emission, and resolved imaging of circumstellar material. In the following sections, we review the observational diagnostics available, the empirical rates and prescriptions derived from them, and the consequences of mass loss for the evolution and ultimate fate of these massive stars.

\subsection{Observational Evidence for Mass Loss}
\label{sec:obsmassloss}
Mass loss from RSGs is inferred through a variety of observational diagnostics, each probing different components of the circumstellar environment and tracing material at different distances from the star. One of the most widely used diagnostics is the infrared excess in the spectral energy distribution (SED) of RSGs. Cool, dust-rich winds produce thermal emission at wavelengths beyond 5--10\,$\mu$m, where the stellar photosphere contribution has fallen well below the dust emission. By fitting the observed SED with models that include both the stellar photosphere and a dusty wind, one can constrain the dust mass-loss rate and, with assumptions about the gas-to-dust ratio, the total mass-loss rate. This technique has been applied extensively to Galactic and LMC/SMC RSGs using data from the \textit{Spitzer Space Telescope} and the \textit{Wide-field Infrared Survey Explorer} (WISE), revealing infrared excesses in a significant fraction of the RSG population \citep{Verhoelst2009}.

While the SED technique is widely used, the mass-loss rates determined depend heavily on the modelling assumptions. In particular, studies differ in how they treat the wind when converting a fitted dust optical depth from a radiative transfer code \citep[such as DUSTY,][]{ivezic1999dusty} into a mass-loss rate, and for the same photometry these choices can change the inferred rate by orders of magnitude. Some analyses treat the dust as radiatively driving the outflow beyond the condensation radius, using the radiatively driven wind solution in DUSTY \citep[e.g.][]{yang2023}, while others assume the dust is carried in a wind of fixed expansion velocity \citep[e.g.][]{Beasor2020, Antoniadis2024}. Radiatively driven wind models return rates two to three orders of magnitude higher than fixed-velocity models for the same stars \citep{Antoniadis2024}, but RSG winds are not thought to be radiatively driven from the photosphere, and the mechanism that launches them remains unknown (Section~\ref{sec:massloss}), so there is little reason to impose a radiatively driven velocity structure. Another key assumption is the adopted grain size. Grains measured around nearby RSGs such as VY~CMa are typically a few tenths of a micron \citep{smith2001vycma, scicluna2015}, larger than the sub-tenth-micron grains of the standard ISM grain size distribution \citep{mathis1977}. Adopting small ISM-like grains in radiatively driven or optical-depth-constrained models inflates the derived $\dot{M}$, whereas in fixed-velocity mid-infrared SED fitting (where grains are in the Rayleigh regime), varying the grain size between 0.1–0.5 $\mu$m leaves the resulting mass-loss rate constant \citep{Beasor2016}. As noted above, the conversion from dust to total mass-loss rate also requires an assumed gas-to-dust ratio, usually a single fixed value for a given metallicity. However, in reality the ratio may change as the star evolves \citep{Beasor2020}. CO-based gas rates compared with SED-based dust rates for nearby Galactic RSGs imply ratios ranging from $\sim$20 to $\sim$500 \citep{Decin2024}, so a fixed value can bias the inferred $\dot{M}$ in either direction. The adopted expansion velocity matters too. Measured RSG wind speeds span roughly 10--30\,\kms\ \citep[e.g.][]{Goldman2017, Decin2024}, so a single assumed value can be wrong by a factor of $\sim$2 for any given star. \citet{Decin2024} showed that their CO-based rates for RSGC1 are lower than the SED-based rates of \citet{Beasor2020} mainly because the latter assumed 25\,\kms, roughly twice the velocities measured from the CO lines, and that the two agree once this is corrected. At least some of the spread between the prescriptions compared in Fig.~\ref{fig:mdot_prescs} therefore reflects modeling assumptions rather than genuine differences between the samples.

Molecular emission lines provide another powerful diagnostic of mass loss. Molecules such as CO, SiO, OH, and H$_2$O form in the cool, dense circumstellar envelopes of RSGs and produce characteristic rotational and ro-vibrational emission lines at infrared and radio wavelengths. CO emission, in particular, has been widely used to trace RSG mass loss, as the CO molecule is abundant, stable, and easily detected in Galactic RSGs and, for the brightest objects, in the Magellanic Clouds. Observations of CO rotational lines at millimeter wavelengths allow direct measurement of the expansion velocity and kinematic structure of the wind, while the line flux provides a measure of the CO column density. With assumptions about the CO abundance and excitation conditions, the mass-loss rate can be derived \citep{Mauron2011,Decin2006}. However, uncertainties in the CO photodissociation rate, the gas kinetic temperature, and the fraction of carbon locked in CO versus other molecules introduce systematic uncertainties of factors of a few in the derived mass-loss rates.

Maser emission provides a complementary probe of the circumstellar envelope, with different species tracing different radii. SiO masers arise within a few stellar radii and probe the wind acceleration zone; H$_2$O masers trace intermediate radii; and OH masers at 1612\,MHz arise in the outer envelope, well beyond the dust formation region, and are commonly detected in RSGs with significant mass loss \citep{Goldman2017}. The brightness and compactness of maser emission allows Very Long Baseline Interferometry (VLBI) observations to spatially resolve the masing regions, providing direct constraints on the size and kinematics of the circumstellar envelope on sub-milliarcsecond scales. Masers are, however, highly non-linear, beamed, and variable, which makes them excellent kinematic tracers but poor mass tracers, constraining the geometry of the outflow far better than its rate. The envelopes of the most luminous RSGs, notably VY CMa and NML Cyg, host a rich circumstellar chemistry, with more than two dozen molecular species identified. This chemistry, and what it reveals about the physical conditions in the outflows, is reviewed in detail by \citet{ziurys2025review}.

Spatially resolved imaging of circumstellar dust shells provides the most direct view of the material itself. Observations with adaptive optics systems on 8--10\,m class telescopes, or with space-based facilities such as HST, can resolve the innermost dust shells around the nearest RSGs. These images reveal complex, often asymmetric structures, including clumps, arcs, and spirals that likely arise from episodic mass loss, stellar pulsations, or the influence of a binary companion \citep{Kervella2016, Montarges2021, smith2001vycma}. Interferometric observations in the infrared and sub-millimeter, using facilities such as the Very Large Telescope Interferometer (VLTI) and the Atacama Large Millimeter/submillimeter Array (ALMA), have pushed spatial resolution to the point where individual dust clumps and the acceleration region of the wind can be resolved. These observations reveal that RSG mass loss is not a smooth, steady process, but rather highly variable and structured on a range of spatial and temporal scales \citep{Ohnaka2017,Montarges2019}. The best-studied example is VY CMa, whose ejecta comprise arcs, knots and clumps of different ages. Proper motions and radial velocities show these to be the products of discrete, directed outflows over the past $\sim$1000 yr rather than a spherical wind \citep{smith2001vycma, humphreys2005, humphreys2007, humphreys2021}. ALMA CO imaging has revealed further large-scale outflows not visible in the optical or infrared \citep{singh2023, humphreys2024, humphreys2025}. We note however that only a handful of objects have been imaged in this way, and all are nearby and atypically bright. More importantly, resolved asymmetry does not by itself imply that the time-averaged mass-loss rate differs from the smooth-wind value --- which is the quantity stellar evolution models require.

The interpretation of these diverse observational diagnostics is complicated by the fact that different tracers probe different physical regimes and may be sensitive to different mass-loss events in the star's history. For instance, infrared dust emission traces the integrated dust mass in the envelope, which depends on the mass-loss history over the dust grain survival timescale, typically $\sim$10$^3$--10$^4$ years. By contrast, molecular line emission traces the current gas density and kinematics, and masers probe even more localised regions close to the photosphere. As emphasised by \citet{vanLoon2025}, a common problem in methods that use tracers of mass loss is the failure to correct for varying scaling factors --- often because there is no information available on which to base such corrections. A notable example is the scaling of dust-mass to mass-loss rate, which requires an assumption of the gas-to-dust ratio, $r_{\rm gd}$. Typically, studies base this assumption off the known $r_{\rm gd}$ of the local environment, so for RSGs in the Milky Way a value of $\sim$200 is assumed, while for the metal-poor LMC a higher value of $\sim$500 is assumed. A blanket $r_{\rm gd}$ for all RSGs may not be a good assumption in the first place, as it is likely the ratio changes throughout the evolution of these systems.

\subsection{Mass-loss rate prescriptions}
\label{sec:prescriptions}

Empirically determined mass-loss rates for RSGs show substantial scatter at fixed luminosity. This substantial star-to-star scatter, even among objects with similar luminosities and spectral types, likely reflects a combination of intrinsic variability in the mass-loss process, differences in evolutionary state and internal structure, and systematic uncertainties in the observational techniques used to derive mass-loss rates. Early compilations of RSG mass-loss rates, such as those presented by \citet{Mauron2011}, revealed this scatter but its physical origin remains unconfirmed.

One of the earliest and most influential mass-loss rate prescriptions was presented by \citet{deJager1988}. In this work, the authors collated as many mass-loss and luminosity measurements as possible for stars of all spectral types and used this heterogeneous dataset as the basis for a new, unified mass-loss rate prescription. The de Jager prescription expresses the mass-loss rate as a function of stellar luminosity and effective temperature, with different functional forms for different regions of the Hertzsprung--Russell diagram. The relation is approximated by:
\begin{equation}
\log(\Mdot) = a \logL + b \log(\Teff) + c,
\end{equation}
where $a$, $b$, and $c$ are empirically determined coefficients. While the de Jager prescription has been widely used in stellar evolution models for decades, particularly for cool supergiants ($\Teff < $ 10,000\,K), it suffers from several limitations. First, it is based on a relatively small and heterogeneous sample of stars (only 15 RSGs), with mass-loss rates derived using a variety of techniques with different systematic uncertainties. The sample is also likely biased toward stars with unusually high mass-loss rates, since these dustier RSGs would have been easier to detect with the ground-based bolometric techniques available at the time, predating the advent of modern space-based infrared astronomy. Since the prescription was released, many of the cool supergiant stars within it have had their luminosities and/or mass-loss rates revised, in some cases by orders of magnitude. Second, it does not explicitly account for metallicity, which may be expected to influence mass loss through its effect on dust opacity and thus radiation pressure. \citet{Beasor2020} showed that the de Jager prescription can overestimate the mass-loss rate at a given luminosity by up to a factor of 30, with the highest luminosity (and hence, highest initial mass) stars being the worst affected.

Subsequent work has sought to improve upon the de Jager prescription by using more homogeneous datasets and incorporating additional physical dependencies. \citet{vanLoon2005} derived mass-loss rates for a sample of dust-enshrouded evolved stars, predominantly AGB stars but including RSGs, in the Large Magellanic Cloud (LMC) using infrared photometry from ISO and ground-based facilities, allowing for a direct comparison between Galactic and LMC stars at different metallicities. While dust formation is expected to be suppressed at lower metallicity, the observed mass-loss rates in the LMC showed only a weak metallicity dependence. This suggests that the total mass-loss rate may be relatively insensitive to metallicity, even if the mechanism by which that mass loss occurs (e.g., dust-driven versus pulsation-driven) varies \citep{vanLoon2025}. 

\begin{figure*}
    \centering
    \includegraphics[width=0.8\linewidth]{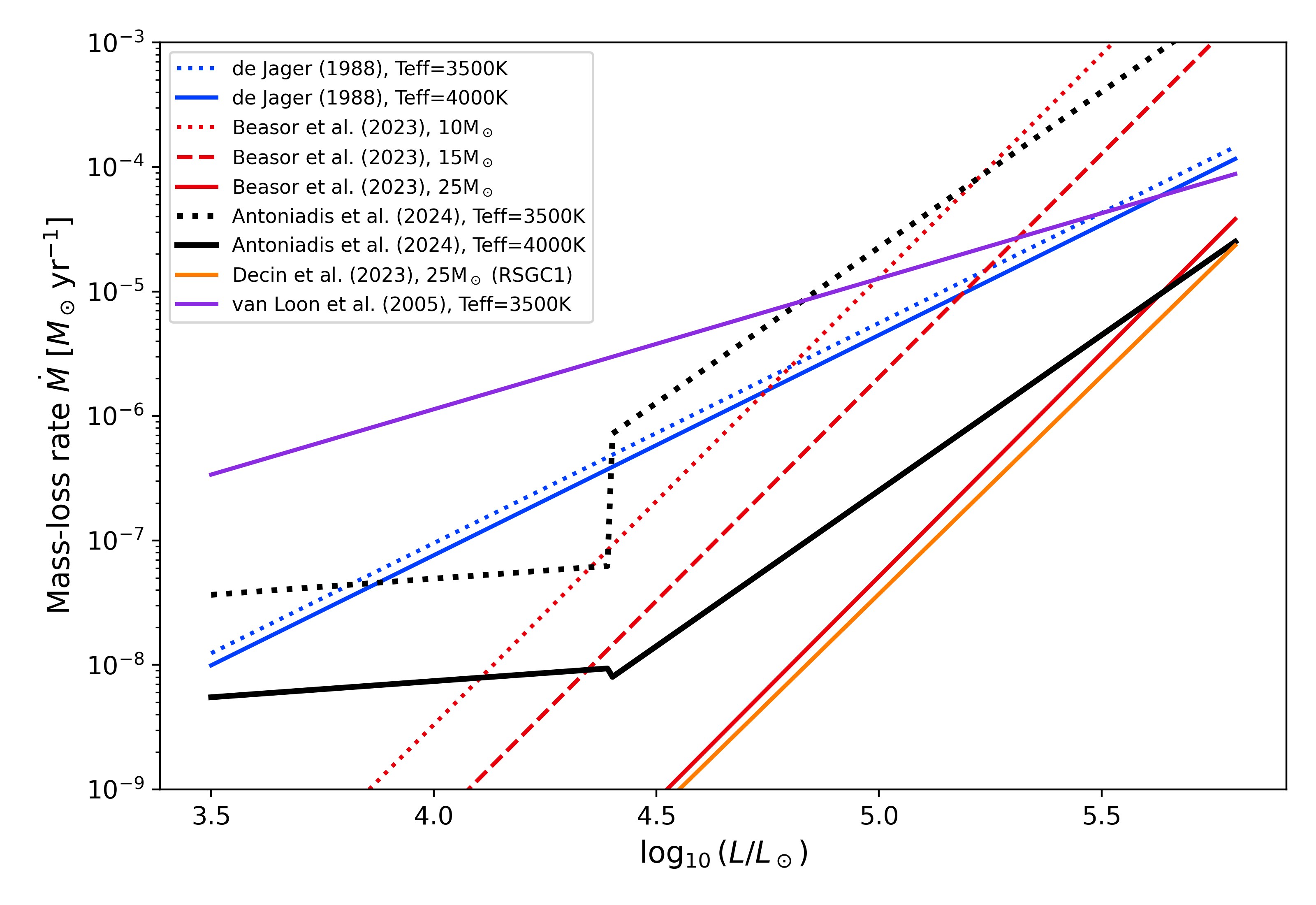}
    \caption{Comparison of RSG mass-loss rate prescriptions as a function of luminosity. For relations depending on effective temperature (\citealt{deJager1988}; \citealt{Antoniadis2024}; \citealt{vanLoon2005}), curves are shown for \Teff\ = 3500\,K (dotted) and 4000\,K (solid), spanning the typical RSG temperature range, except for \citet{vanLoon2005} where only \Teff\ = 3500\,K is shown. For relations depending on initial mass, \citet{Beasor2020} is shown for $M_{\rm ini}$ = 10, 15, and 25\,\Msun\ (dotted, dashed, solid), spanning their calibration sample, while \citet{Decin2024} is shown only for $M_{\rm ini}$ = 25\,\Msun, matching the initial mass of RSGC1 (the cluster targeted by their CO-based measurements). The \citet{Antoniadis2024} relation shows a pronounced kink at $\log(L/\mathrm{L}_\odot) \approx 4.4$ and a correspondingly strong sensitivity to the assumed \Teff, particularly above the kink (see main text).}
\label{fig:mdot_prescs}

\end{figure*}

More recent prescriptions have used multi-wavelength SEDs (such as \citet{Beasor2020} and \citet{Antoniadis2024}) or ALMA CO rotational line observations (\citet{Decin2024}) to derive mass-loss rates for RSG populations in the Galaxy and the Magellanic Clouds, see Fig.~\ref{fig:mdot_prescs}\footnote{For clarity, only a subset of the available prescriptions is shown. Among those not shown, \citet{yang2023} derived rates for $\sim$2000 RSGs in the SMC assuming a radiatively driven wind (Section~\ref{sec:obsmassloss}), and consequently find substantially higher rates than those plotted here, particularly for the most luminous stars. As noted in Section~\ref{sec:massloss}, the winds of RSGs are unlikely to be radiatively driven.}. Both \citep{Beasor2020} and \citep{Decin2024} target RSGs in clusters. Targeting objects in a cluster, assumed to be a coeval population, has a number of advantages. For one, clusters typically have well-determined distances, meaning the estimated RSG luminosities will be more reliable. In addition, a coeval population has a known age (particularly when anchored to the lowest-luminosity RSG turnoff to mitigate contamination from binary mass gainers and merger products, see e.g. \citep{Beasor2019, Britavskiy2019}, and discussion in Section~\ref{sec:binarity}), and therefore a known initial mass for the single stars currently in the RSG phase, so mass-loss rate can be measured as a function of evolutionary stage at fixed mass and metallicity. Given the short lifetime of RSGs ($\leq$ 10$^6$ years), single RSGs in a given cluster share the same initial mass to within ${\sim}1$--$2\,\Msun$. This removes the degeneracy between initial mass and evolutionary state that affects field samples. The downside to focusing on RSGs in clusters is sample size, with only a handful of RSG-rich clusters available for this work. However, a single cluster can still exceed the size of the entire de Jager sample (15 RSGs); NGC~2100 alone contains 20 \citep[see][]{Beasor2020}. Combining several clusters, \citet{Beasor2020} found that $\dot{M}$ depends on both $L$ and $M_{\rm ini}$, so that at a given luminosity, stars of higher initial mass have lower mass-loss rates (see red lines in Fig. \ref{fig:mdot_prescs}). This has consequences for how the measurements are fitted. \citet{humphreys2020} analysed RSGs in Galactic clusters, several of which are in common with \citet{Beasor2020}, but fit $\dot{M}$ as a function of luminosity alone, and report a transition to higher rates above $\sim$$10^5\,L_\odot$. Within each cluster $\dot{M}$ rises steeply with $L$, but at lower luminosities several clusters of different initial mass overlap, and since the higher-mass clusters sit at lower $\dot{M}$ for a given $L$, a single fit across them is flattened. At the highest luminosities the sample is dominated by a single cluster (RSGC1), so the fit recovers the steep slope of an individual cluster, producing an apparent upturn. When the clusters are considered individually, the \citet{humphreys2020} results show the same dependence on initial mass as found by \citet{Beasor2020}, and once this is accounted for no upturn is required.

Other studies, such as \citet{Antoniadis2024}, instead use a large sample of RSGs across a single galaxy. This gives better statistics and a wider luminosity range, but the initial masses of individual stars are not known, so the scatter introduced by mixed masses cannot be removed in the same way. The \citet{Antoniadis2024} relation shows a turning point at $\log(L/L_\odot) \approx 4.4$, above which the rates rise steeply (Fig.~\ref{fig:mdot_prescs}). Above this kink the relation also depends strongly on the effective temperature, where a 500\,K change shifts the predicted rate by a factor of ${\sim}100$; below the kink the same shift changes the rate by a smaller factor of ${\sim}7$. \citet{Antoniadis2025} note that this steep temperature dependence is driven largely by the adopted colour--\Teff\ calibration rather than a robustly determined physical effect, and caution against extrapolating it beyond ${\sim}4500\,$K. Both approaches find typical mass-loss rates of ${\sim}10^{-6}$ to $10^{-5}\,\Myrunit$. Both also find that the mass-loss rates implemented in many evolutionary models overestimate the total mass lost during the RSG phase.

If steady RSG winds are as weak as these measurements suggest, they cannot remove enough of the envelope to change how a star ends its life, and so single star evolution cannot explain populations of stripped stars. In single-star evolution, an RSG that loses most of its hydrogen envelope evolves back to the blue, producing yellow supergiants, Wolf--Rayet stars and stripped-envelope SNe without invoking a binary companion \citep[e.g.][]{Georgy2012, Meynet2015}, but weak winds rule out the single star pathway for producing stripped stars \citep{beasor2022dersgs}. Several observations have been suggested to imply that some RSGs lose substantial mass, including the apparent lack of high-mass Type~II-P progenitors (Section~\ref{sec:rsgproblem}), yellow supergiant progenitors of Type~IIb SNe, and dense CSM around many SNe at the point of explosion. This has motivated the suggestion that much of the mass is lost in short-lived eruptive episodes, which would be largely missed by measurements of the time-averaged rate \citep[e.g.][]{Cheng2024, bonanos2024episodic}.

 The eruptive scenario carries a testable consequence. If episodic events remove several \Msun{} over the RSG phase, there should be a population of RSGs caught during or shortly after such an episode, and these would be heavily dust obscured with large infrared excesses. \citet{beasor2022dersgs} tested this by modelling the mid-IR excess of the full LMC RSG sample. Of the 14 objects previously proposed as dust enshrouded, only one, WOH~G64, has both a high mass-loss rate ($\dot{M} \geq 10^{-4}$\,\Myrunit) and high circumstellar extinction ($A_{\rm V} = 7.9$\,mag)\footnote{It is important to note that WOH~G64 is {\it not} a typical RSG. Its TiO bands disappeared for several years and its spectrum temporarily resembled that of a yellow hypergiant \citep{munoz2026}, but by 2024--2025 the TiO bands had returned \citep[][Shaqil et al. (subm.)]{vanloon2026woh}. A transition to warmer temperatures and back in less than a decade is far too rapid to be explained by enhanced mass loss from a single star, even with mass-loss rates ten times the \citet{deJager1988} prescription, evolutionary models take thousands of years to move an RSG to the yellow (Shaqil et al. subm.). Together with composite spectral features that indicate a hotter companion, this suggests that WOH~G64 is a binary, in which a close interaction stripped part of the outer envelope, which was then rebuilt on a thermal timescale of a few years \citep[][Shaqil et al. (subm.)]{munoz2026}. Excursions between the RSG and yellow hypergiant states are rare, and are seen only among the most luminous cool supergiants close to the H--D limit. The best-known previous example is Var~A in M33, which showed an M-type spectrum with strong TiO bands and heavy circumstellar reddening in the 1980s \citep{humphreys1987vara} before returning to a warmer yellow hypergiant state, a change attributed to a high mass-loss episode lasting several decades \citep{humphreys2006vara}.}. The rest are either normal RSGs with weak infrared excess or lower mass AGB stars. If a heavily obscured phase is one that all RSGs pass through, it lasts less than $3 \times 10^{4}$\,yr and removes at most 1--2\,\Msun. This is not enough to strip the envelope, and it rules out the single star channel for producing stripped stars by winds below roughly 40\,\Msun. 

This constrains sustained high mass-loss phases, but not brief ones. An event lasting months rather than millennia would leave no lasting population of obscured stars and would not be detected this way. Recent theoretical work predicts episodic mass loss of exactly this kind, though the two current models differ substantially in what they require. \citet{Cheng2024} implement an energy-based eruptive mass-loss scheme in MESA, in which material above regions of locally super-Eddington radiative luminosity is ejected when the excess energy exceeds the binding energy of the overlying envelope. Mass-loss rates reach $\sim$$10^{-2}$\,\Myrunit{} in a band between $3.5 < \log(\Teff/\mathrm{K}) < 4.0$. At full efficiency, the mechanism prevents stars above 20\,\Msun{} from becoming RSGs at all, which the authors suggest as an explanation for the lack of RSGs above 20--30\,\Msun. The efficiency is a free parameter, however, and at a tenth of that value all their models still become RSGs. \citet{Fuller2024} take a different approach, arguing that shocks driven by convection support a dense chromosphere out to the dust formation radius at $\sim$5\,$R_*$. Mass loss is dominated by uncommon convective episodes with velocities around three times the typical value, which for a Betelgeuse-like star implies a major ejection every 30--100\,yr, each lasting a convective turnover time and removing $\sim$$10^{-4}$\,\Msun. The ejected dust reaches optical depths of order unity, briefly obscuring the star.

These predictions are not equivalent. The events described by \citet{Fuller2024} are frequent and small, and are consistent with the constraint above, since they neither remove significant mass nor produce a lasting population of obscured stars. Those of \citet{Cheng2024} are large enough to alter the evolutionary track, and the corresponding population of stripped or partially stripped objects should be identifiable. Whether either operates across the RSG population is not yet known, although individual episodic events have now been observed in a few nearby, luminous stars. For example, the 2019--2020 Great Dimming of Betelgeuse followed a dense outflow \citep{dupree2020, dupree2022} and has been attributed to dust condensing in the ejected gas \citep{Montarges2021}, although it can also be explained by changes in the wind without dust \citep{daviesplez2021}, while RW~Cep underwent a similar dimming in 2022 \citep{anugu2023rwcep, anugu2024rwcep} and the ejecta of VY~CMa record repeated discrete outflows that may dominate its mass loss \citep[e.g.][]{smith2001vycma,smith2009vycma, humphreysjones2022, humphreys2024}. These are, however, among the nearest and most luminous cool supergiants, and how often such events occur in typical RSGs, how much mass they remove (i.e., if there is enough mass lost to affect onward evolution), and the mechanism driving these events, remains unknown.

What is needed is a measurement of how often such events occur across the population. A mass ejection of the magnitude predicted by \citet{Fuller2024} would produce a dimming event similar to that of Betelgeuse, on a timescale of months, and would be detectable photometrically. Long baseline monitoring of large RSG samples is therefore the most direct test available. \citet{lauriano2026} have assembled GalRSG, a catalogue of 227 Galactic RSGs in the Scutum--Crux region at distances of 4--7\,kpc, and are monitoring it in a five-year campaign begun in 2023, sampling field stars every $5 \pm 2$ days with the VST and cluster members every two days with REM in the optical and near-IR. The sample is largely coeval and co-distant, and spans initial masses of 8--15\,\Msun, precisely the range where extragalactic transient surveys are least sensitive. Campaigns of this kind should either detect eruptions at the rate the models predict or place a useful upper limit on their frequency. Until they do, the frequency of eruptions large enough to change a star's evolution will remain unconstrained.

A final opportunity for a star to lose mass comes in the years prior to supernova. There is growing evidence that mass-loss rates may increase dramatically in the final years to decades before core collapse, as discussed in the context of circumstellar interaction in Section~\ref{sec:csm}. If such enhanced pre-supernova mass loss is common, the mass-loss rates derived from long-lived circumstellar envelopes (which trace mass loss averaged over $\sim$$10^3$--$10^4$ years) may significantly underestimate the peak rates achieved just before explosion. However, given the likely short duration of these pre-SN mass-loss episodes -- of order a few hundred years for the more extreme cases inferred for Type~IIn SNe, and considerably shorter, down to just a few years, for the early interaction signatures seen in some Type~II-P events -- it is unlikely that this would alter the evolutionary path of a star. Instead, it is more likely to cause the star to explode into a dense circumstellar medium (CSM), and hence appear as a different type of SN (Type~IIn rather than Type~II-P).

\subsection{Physical Mechanisms}

Mass loss in RSGs is generally understood as a multi-step process, because no single mechanism operates across the whole range of radii. Radiation pressure on spectral lines, which drives the winds of hot stars, is ineffective in a cool atmosphere where most metals are bound in molecules; radiation pressure on dust grains is effective, but only beyond the condensation radius at several stellar radii. Material must therefore first be levitated from the photosphere to that radius by some other means. Vigorous envelope convection, which in RSGs extends nearly to the photosphere, generates large-scale flows and acoustic waves that steepen into shocks and deposit momentum in the outer atmosphere \citep{Freytag2002, Chiavassa2011}; large-amplitude pulsations, with periods of hundreds to thousands of days, periodically extend the atmosphere and reduce the local effective gravity \citep{Hofner2003, Bladh2019}; and turbulent pressure in the extended atmosphere has recently been proposed as a further contributor \citep{Kee2021}.

Once material reaches temperatures below $\sim$1000--1500\,K, silicate and alumina grains condense. These have large cross-sections at the near- and mid-infrared wavelengths where RSGs emit most of their radiation, so radiation pressure accelerates them outward, and collisional coupling drags the gas with them to the observed terminal velocities of 10--30\,\kms. The efficiency of this final step depends on the grain size distribution, composition, and dust-to-gas ratio, all of which vary with metallicity and with the local chemistry of the wind --- which is one reason why a prescription calibrated in one environment need not transfer to another.

\subsection{Consequences for Stellar Evolution}

The mass-loss rate during the RSG phase has significant consequences for the subsequent evolution of massive stars and for the properties of the supernovae they produce. The total mass lost during the RSG phase determines the final hydrogen envelope mass at the time of core collapse, which in turn dictates the supernova type and strongly influences the light curve properties. Stars that lose only a modest amount of mass ($\lesssim$1--2\,\Msun) during the RSG phase will retain hydrogen-rich envelopes and produce Type II-P or II-L supernovae, although the distinction between these two classes is contested and the observed decline rates may form a continuum \citep{Anderson2014}, characterized by prominent hydrogen lines in their spectra and, in the case of II-P, a long-lasting plateau in the optical light curve. By contrast, in single-star models stars that experience more extreme mass loss may shed the bulk of their hydrogen envelopes and evolve to hotter temperatures, becoming yellow or blue supergiants before core collapse \citep[e.g.][]{Yoon2010, Georgy2012, Meynet2015}. Such stars are thought to be the progenitors of Type IIb supernovae, which show hydrogen in their early spectra but transition to hydrogen-poor spectra at later times. If mass loss is severe enough to remove the hydrogen envelope entirely, the star may produce a Type Ib or Ic supernova, depending on whether the helium layer is also stripped \citep{Smith2014}. We note the single-star wind channel is not the favoured explanation for most stripped-envelope SNe: the observed rates are difficult to reproduce with winds alone, and binary stripping is now generally preferred \citep{smith2011rates, Eldridge2008}. This is supported by the low ejecta masses inferred for most stripped-envelope SNe, which are difficult to reconcile with the massive single stars that would be needed to strip their envelopes by winds \citep{drout2011, eldridge2013, lyman2016}, and by the detection of a binary companion to the progenitor of the Type~IIb SN~1993J \citep{maund2004, fox2014,zapartas2026binarysn}.

Beyond determining the supernova type, mass loss affects the surface abundances of the progenitor star, with potential implications for chemical yields and galactic chemical evolution. As material is lost from the surface, the composition of the wind reflects the processed material dredged up from the interior by convection. For RSGs, this means that the wind is enriched in products of hydrogen and helium burning, including $^{14}$N, $^{13}$C, and $^{17}$O. The total mass returned is, however, modest: on the revised rates discussed above, a star of $\sim$15\,\Msun{} sheds well under 1\,\Msun{} over the whole RSG phase, so the contribution to galactic nitrogen enrichment is correspondingly smaller than older prescriptions imply. The degree of enrichment depends on the amount of mixing between the convective envelope and the hydrogen-burning shell, which in turn depends on the stellar mass, rotation, and evolutionary state.

Mass loss also shapes the circumstellar medium encountered by the eventual explosion, with consequences for the light curve and spectra of the resulting supernova; this is discussed in Section~\ref{sec:csm}.

Finally, uncertainties in RSG mass-loss rates propagate through stellar evolution models and affect predictions for a wide range of astrophysical phenomena. For instance, the number ratio of red to blue supergiants in a stellar population depends on the duration of the RSG phase, which is sensitive to the mass-loss rate \citep{Davies2018}. Similarly, the final masses of black holes and neutron stars produced by core collapse depend on the envelope mass at explosion, and thus on the integrated mass loss over the star's lifetime.

\section{Red supergiants as supernova progenitors}\label{sec:progenitors}

Core-collapse supernovae (CCSNe) represent the final stage of evolution for massive stars with initial masses above $\sim$8\,\Msun, and RSGs have long been expected to be, and are now confirmed as, the immediate progenitor for the majority of hydrogen-rich CCSNe, Type II-P. This connection between RSGs and Type II-P SNe arises from stellar evolution theory, since RSGs have large convective envelopes and can reach large radii, resulting in the characteristic plateau phase seen for II-P. During the plateau, the photosphere tracks the hydrogen recombination front as it moves inward in mass coordinate through the expanding ejecta. Because recombination fixes the temperature at which the ejecta become transparent, the photospheric radius and temperature both remain approximately constant, and the luminosity is nearly flat for 50--100 days following the explosion \citep[e.g.][]{goldberg2020}. It is this mechanism that makes the plateau a signature of a massive hydrogen envelope, and hence of an RSG progenitor. Over the past two decades, direct observational evidence has confirmed this link between RSGs and hydrogen-rich SNe through the identification of progenitor stars in serendipitous pre-explosion archival images, mostly taken by the \textit{Hubble Space Telescope}. At the same time, these progenitor studies have revealed several important puzzles regarding the final stages of massive star evolution. In particular, the claimed lack of high-mass RSG progenitors among observed SN populations has instigated debates about single-star stellar evolutionary theory, raising questions about mass-loss rates, stellar winds, and the role of dust obscuration in the final years before core collapse.

The most direct method of linking SNe to their progenitors is the identification of a candidate progenitor in such images via precise astrometric alignment with post-explosion imaging \citep{Smartt2009,VanDyk2017}. Progenitor identification typically proceeds in several stages. First, the precise location of the SN must be established through high-resolution imaging of the SN itself, either with HST or ground-based adaptive optics systems. This position is then transferred to pre-explosion images through astrometric alignment, accounting for distortions, proper motions, and systematic uncertainties. Once a candidate progenitor is identified, the available photometry --- in practice usually one or two bands --- is converted to a bolometric luminosity using an assumed bolometric correction, and that luminosity compared with the terminal luminosities of stellar evolution models to estimate an initial mass.

The first systematic efforts to identify RSG progenitors began in the early 2000s, with the identification of the progenitor of SN 2003gd \citep{VanDyk2003, Smartt2004}. This discovery utilised a combination of HST imaging at optical wavelengths and ground-based near-infrared observations to characterise the progenitor star. Since then, the number of confirmed RSG progenitors has grown steadily, aided by the increasing depth and wavelength coverage of archival surveys. The \textit{Spitzer Space Telescope} proved particularly valuable for characterising RSG progenitors, as these cool stars emit the majority of their radiation in the infrared. 
More recently, the \textit{James Webb Space Telescope} (JWST) has begun to play a role, enabling high-precision post-explosion astrometric alignment to identify the progenitor of SN 2022acko in archival HST images \citep{vandyk2023acko} and directly capturing the dusty progenitor of SN 2025pht in pre-explosion imaging shortly before explosion \citep{kilpatrick2025pht}. JWST changes the analysis qualitatively rather than incrementally: it is the first facility able to measure the full SED of a dusty progenitor, so the bolometric luminosity can be measured rather than extrapolated from one or two optical bands. This bears directly on the dust-obscuration explanation of the Red Supergiant Problem discussed below. To date, approximately two dozen Type II-P SNe have confirmed RSG progenitor detections \citep{VanDyk2025}. New detections currently accrue at a rate of roughly one per year, almost entirely from archival HST imaging. Archival data from the Euclid and CSST wide-field surveys could raise this to $\sim$10--20 per year for progenitors of 8--25\,\Msun{}, although \citet{wu2026euclid} regard this estimate as optimistic.
The identified progenitors span a range of inferred luminosities, with $\logL \sim 4.0$--5.2, corresponding to initial masses of approximately 8--18\,\Msun (the upper limit in particular is sensitive to the chosen initial mass-terminal luminosity relation used). Effective temperatures of 3000--4000\,K and radii of several hundred to over 1000\,\Rsun{} are typically \emph{assumed} in the SED fitting rather than measured, since the photometry is rarely sufficient to constrain them. The SEDs of these progenitors can generally be fit with cool photospheric models, though the more luminous progenitors often require significant amounts of circumstellar dust to reproduce their observed infrared excess. This dust suggests the presence of circumstellar material ejected during the RSG phase, with important implications for the interpretation of progenitor masses and the nature of pre-SN mass loss.

Another important factor to understand when determining progenitor masses is that this is a process fraught with uncertainties and biases at every step. Typically, a progenitor will only have been observed prior to explosion in one or two wavebands, usually \textit{I}-band and \textit{V}-band ($\lambda \simeq$ 0.81 and 0.55\,$\mu$m, respectively), meaning that assumptions have to be made to convert the fluxes to a bolometric luminosity. The first is the extinction towards the progenitor, typically estimated either from the reddening of neighbouring stars \citep[e.g.][]{maund2005}, on the assumption that the progenitor was reddened by a similar amount, or from Na\,\textsc{i}\,D absorption measured in the SN spectrum itself around peak light. This takes into account foreground extinction and the extinction within the host association itself, but neither method captures circumstellar extinction local to the progenitor which can contribute significantly to the total extinction (see discussion below) since any close-in CSM is destroyed by the shock breakout before peak light. Next, the extinction-corrected photometric flux can be converted to a bolometric luminosity via a bolometric correction. In the original red supergiant problem paper, \citet{Smartt2009} applied a uniform bolometric correction to all progenitors, assuming they were all early-type M stars. From here the luminosities can be converted to initial masses via terminal luminosity--initial mass relations presented in stellar evolution models \citep[see Fig.~1 within][]{Smartt2009}. Depending on the model chosen, there may be a systematic offset of the initial mass by up to 2\,\Msun.

These systematics do not act independently, and their combined effect is what matters for the argument of Section~\ref{sec:rsgproblem}. Extinction, bolometric correction, distance, and the adopted luminosity--initial mass relation each contribute, and a realistic total uncertainty on a single progenitor mass is of order $\pm$5\,\Msun. Set against a claimed discrepancy between an observed cutoff at 18\,\Msun{} and a predicted one at 25--30\,\Msun, an uncertainty of this size accounts for a substantial fraction of the gap, even before any physical explanation is invoked.

\begin{figure*}
    \centering
    \includegraphics[width=0.8\linewidth]{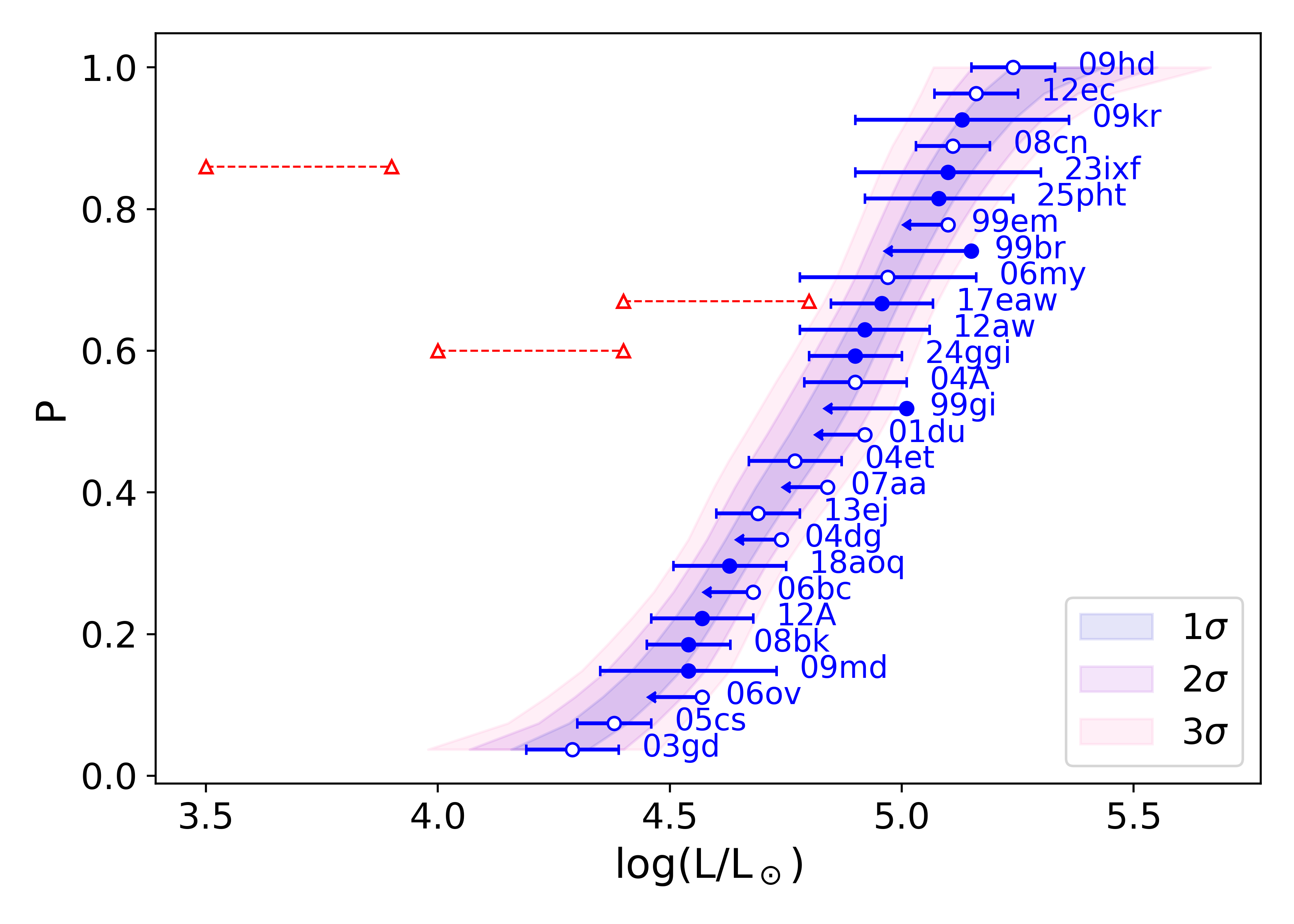}
    \caption{Luminosity distribution of Type II-P SN progenitors detected in pre-explosion imaging, adapted from \citet{beasor2025} to include SN~2025pht. The distribution is consistent with there being no missing RSGs to within 3$\sigma$. Red triangles denote progenitors with pre-explosion near- and/or mid-IR data, and show the luminosity that would be inferred for each progenitor in the absence of mid-IR data, using standard bolometric correction (BC) assumptions. For SN~2025pht, the BC method would drastically underestimate the total luminosity; its resulting value lies off the scale of the plot and is not shown.}
    \label{fig:cld}

\end{figure*}

\subsection{The Red Supergiant Problem}
\label{sec:rsgproblem}

One of the most controversial claims to emerge from progenitor studies is the so-called ``red supergiant problem'' --- the apparent absence of detected progenitors above initial masses of $\sim$17--18\,\Msun{} \citep{Smartt2009,Smartt2015}. This upper mass limit has been reproduced in several studies and is significantly lower than theoretical predictions from single-star stellar evolution models. Standard evolutionary models predict that stars with initial masses up to $\sim$25--30\,\Msun{} should retain their hydrogen envelopes and explode as RSGs, depending on rotation, metallicity, and mass-loss prescriptions. Whether the discrepancy between the observed and predicted distributions is statistically significant is, however, disputed, and we return to this below.

The red supergiant problem manifests in two complementary ways. First, the direct detection of SN II-P progenitors through pre-explosion imaging appears to reveal an upper mass cutoff. Despite intensive searches in nearby galaxies with extensive archival coverage, no progenitors with initial masses above $\sim$18\,\Msun{} have been confirmed for normal Type II-P SNe. Second, analysis of nebular-phase spectra of SNe II-P, which probe the oxygen-rich core of the progenitor, also suggests progenitor masses at the lower end of expectations \citep{Jerkstrand2014}. The nebular-phase emission line ratios, particularly those involving [O\,{\sc i}], depend sensitively on the oxygen mass synthesised during the star's lifetime, and thus provide a partly independent constraint on progenitor mass. These measurements consistently point to progenitor masses below $\sim$15--18\,\Msun, in broad agreement with the pre-explosion imaging results. The independence is only partial, however: the nebular masses are derived from the same family of stellar evolution models that supply the luminosity--initial mass relation used in the imaging analysis, so a systematic in those models would displace both methods in the same direction. The apparent agreement of two techniques is a large part of why the problem is regarded as robust, and it is weaker evidence than it appears. A third route --- hydrodynamic modelling of the plateau light curve --- has in places favoured somewhat higher masses \citep{Martinez2022}.

Several potential solutions to the red supergiant problem have been proposed, each with different implications for massive star physics. One possibility is that the most massive RSGs are simply missing from photometric samples because they do not explode as Type~II-P SNe at all: with the highest mass-loss rates late in life, they may instead explode as Type~IIn SNe \citep{smith2009vycma}, whose progenitors are obscured by dense circumstellar dust. \citet{Smartt2009} and others have argued that the known rate of Type~IIn SNe is too low to account for the missing high-luminosity RSGs, but these estimates were based on reported discovery rates rather than volume-limited rates corrected for observational selection effects, and so may significantly underestimate the true Type~IIn fraction \citep{smith2011rates}. This same circumstellar dust could also cause pre-explosion imaging studies to underestimate progenitor luminosities more generally \citep{Walmswell2012}: if the most massive RSGs are surrounded by optically thick dust shells in the years before explosion, their true luminosities might be hidden, leading to systematically low mass estimates.

A second explanation invokes enhanced mass loss during the RSG phase for the most massive stars. If RSGs with initial masses above $\sim$18\,\Msun{} experience dramatically increased mass-loss rates in their final years, they could shed their hydrogen envelopes and evolve to warmer temperatures before exploding. Such stars might then appear as yellow supergiants (YSGs) or even blue supergiants (BSGs) at the time of explosion, and would be progenitors of Type IIb or Type Ib/Ic SNe rather than Type II-Ps \citep{Yoon2010,Georgy2012,Meynet2015}. There is some observational support for this scenario, as several Type IIb SNe have been linked to YSG progenitors \citep{Maund2011,VanDyk2014}. The broader possibility that high-mass RSGs preferentially die as other SN flavours (II-L, Ib, etc.) rather than II-P was discussed as viable by \citet{smith2011rates}, based on the relative rates of these SN types. However, the mass-loss mechanism required to drive this evolution remains uncertain, and single-star models typically require very high, perhaps unphysically large, mass-loss rates to reproduce the YSG progenitors of Type IIb SNe. This explanation also sits awkwardly with the argument of Section~\ref{sec:massloss}. It requires RSG winds to be substantially {\it stronger} than measured, for precisely the stars in question, while recent mass-loss measurements indicate they are weaker than previously assumed, by up to a factor of 30. Signatures of early CSM interaction in a large fraction of otherwise normal SN~Type II-P \citep[e.g.][]{Bruch2023} may suggest very compact dense shells are ejected shortly before core collapse, meaning progenitors may have higher extinction from CSM than the normal RSG population. 

A third, and perhaps the most exotic, possibility is that the most massive RSGs do not explode at all, but instead undergo direct collapse to black holes without producing a bright optical transient --- so-called ``failed supernovae'' \citep{Kochanek2008,Kochanek2014}. In this scenario, stars above some critical mass (or, more accurately, above some critical value of a compactness parameter that depends on the core structure) fail to launch a successful supernova explosion, and instead collapse directly to form a black hole. The neutrino-driven explosion mechanism is known to be sensitive to the structure of the stellar core, particularly the compactness of the iron core and the density profile of the overlying mantle. If the most massive RSGs systematically fail to explode, this would explain the absence of high-mass progenitors. The explanation is less clean than it first appears, however, because compactness may not be monotonic in initial mass --- it varies with the carbon and oxygen burning shell structure \citep{Sukhbold2014, Schneider2024} --- so the failed-SN scenario does not naturally produce the sharp mass cutoff that the observations require. In addition, recent simulations of core collapse in high mass progenitors show that high compactness may not correlate with failed explosions \citep[e.g.][]{burrows2023compactness}. Observational searches for failed supernovae have identified a few candidate events where an RSG appears to vanish without a corresponding bright SN \citep{Adams2017,neustadt2021}, providing tentative support for this scenario. The best-studied candidate, N6946-BH1, has since been resolved by JWST into a luminous infrared source consistent with a surviving dust-enshrouded star or merger remnant rather than a vanished progenitor \citep{Beasor2024}, which weakens that support considerably. However, the expected rate of failed SNe and the precise mass range over which they occur remain highly uncertain. At present, with only a couple of candidates (both of which also appear consistent with mergers), failed SNe do not appear to be a clear solution to the RSG problem.

A fourth explanation is that there is nothing to explain. The initial mass function is steeply declining and so high-mass stars are rare, while the number of nearby galaxies with archival coverage deep enough for a progenitor detection is small. The number of high-mass RSG progenitors expected in a sample of two dozen is therefore of order one, and its absence carries little statistical weight. \citet{Davies2018} and \citet{Davies2020} showed that the observed distribution is consistent with a standard IMF and an upper mass limit well above 18\,\Msun{} once the sample size is treated properly. \citet{beasor2025} tested the standard method of determining progenitor luminosities against the RSGs in M31, which have well sampled SEDs from the optical to the mid-IR, a known distance and foreground extinction. Treating each star as though only single-band \textit{I}-band photometry were available (as is the case for most SNe progenitors), and applying the bolometric correction used in the original analyses, it was found that the inferred luminosity is systematically low by 0.3\,dex, a factor of two, with a further 0.2\,dex of scatter caused by unknown spectral type and variability.  When the additional uncertainty is propagated through the luminosity distribution, even the most luminous known RSGs cannot be excluded from the progenitor sample at the 3$\sigma$ level.

The few progenitors with pre-explosion infrared detections confirm this directly. For SN\,2017eaw, SN\,2023ixf and SN\,2024ggi, the luminosity derived from the full SED is higher than the value that would have been obtained from \textit{I}-band photometry alone, by between 0.5 and 1.6\,dex \citep{beasor2025}. \citet{kilpatrick2025pht} make the same point through the bolometric corrections. For the four progenitors detected beyond 2\,$\mu$m, including SN\,2025pht, the bolometric corrections required are on average 2.8\,mag larger than those adopted for the sample detected only in the optical, corresponding to $1.1 \pm 0.5$\,dex in luminosity. Progenitor luminosities derived without infrared data are therefore likely systematically underestimated. Figure~\ref{fig:cld} shows the current luminosity distribution, which remains consistent with no missing progenitors at the 3$\sigma$ level. 

Binary interaction adds a further systematic to these comparisons. Mergers and mass gainers modify the expected luminosity distribution of RSGs, and break the single-star mapping between progenitor luminosity and initial mass on which the progenitor masses are based \citep[Section~\ref{sec:binarity};][]{Zapartas2019, zapartas2021, wang2025rsgclusters}. Any statistical comparison between the observed progenitor sample and model predictions should therefore account for the binary history of the parent population.

\subsection{Circumstellar Interaction}
\label{sec:csm}

Evidence for dense circumstellar material (CSM) around RSG progenitors has emerged as an important observational signature of enhanced mass loss in the years to decades before explosion. This CSM, detected through its interaction with the expanding SN ejecta, indicates that at least some RSGs experience elevated mass-loss rates shortly before core collapse, well above the steady wind-driven mass loss expected from pulsation-driven winds. The interaction between the SN shock wave and pre-existing CSM produces characteristic observational signatures, including narrow emission lines from photoionized CSM (producing Type IIn spectra), modified light curve evolution, and enhanced luminosity at early times.

The presence of CSM around RSG progenitors can be inferred from several observational diagnostics. The clearest signature is early-time spectra showing narrow emission lines (e.g. H$\alpha$, He II, N III/IV) superimposed on the broad features typical of a Type II SN. These narrow lines, with intrinsic widths of tens of \kms, often accompanied by broader electron-scattering wings, indicate the presence of slowly moving circumstellar gas photoionised by the SN shock breakout flash. The detection of such features requires rapid spectroscopic follow-up, as they typically fade within days to weeks of explosion as the CSM is swept up by the ejecta. Spectroscopy obtained within hours to days of explosion has revealed narrow CSM signatures in a substantial fraction of Type II SNe observed early enough, indicating that some degree of enhanced pre-explosion mass loss is common among RSG progenitors \citep[e.g.][]{niemela1985sn,benetti1994narrow,shivvers2015,Gal-Yam2014,smith2015iptf,Khazov2016,Bruch2023}.

Early-time light curves also provide constraints on CSM properties. The interaction of the SN ejecta with a dense CSM shell can produce a luminosity excess at early times, manifesting as a bright initial peak followed by a decline to the plateau phase \citep[e.g.][]{morozova2017}. The timescale and luminosity of this interaction-powered peak depends on the mass and extent of the CSM, with denser and more extended CSM producing brighter and longer-lasting peaks. Detailed modeling of these early light curves can constrain the mass-loss history of the progenitor in the final years before explosion \citep{morozova2017}. Such studies suggest that some RSG progenitors experience mass-loss rates of $10^{-4}$--$10^{-2}$\,\Myrunit{} in the final years to decades before core collapse, significantly higher than typical steady RSG wind mass-loss rates of $\sim$$10^{-6}$--$10^{-5}$\,\Myrunit. The corresponding CSM masses are nevertheless small, typically $\sim$0.01--0.1\,\Msun, so these episodes reshape the appearance of the supernova without materially altering the star's evolution --- the distinction drawn at the end of Section~\ref{sec:massloss}. The inferred rates are also degenerate with the assumed wind velocity and with the CSM geometry: confinement in a disc or shell rather than a spherical wind changes the derived rate substantially.

The physical mechanism driving this enhanced pre-SN mass loss remains uncertain. One possibility is that vigorous convection in the RSG envelope drives large-amplitude pulsations or eruptions, similar to those observed in luminous blue variables (LBVs) or the great eruption of $\eta$ Carinae. RSGs are known to be long-period variables, with pulsation periods ranging from hundreds to thousands of days, and it has been suggested that particularly energetic pulsations could eject substantial amounts of mass \citep{Yoon2010,Quataert2012}. Alternatively, enhanced mass loss could be triggered by structural changes in the stellar interior as the core approaches collapse, such as shell burning or core neon/oxygen burning \citep[see e.g.][]{smith2014preparing,laplace2025}. A further possibility is that the dense material is not an unbound wind at all, but remains bound to the star as an extended, dense chromosphere supported by pulsations or convectively driven shocks \citep{Fuller2024, laplace2025, ma2025arepo}. In this case the early interaction need not imply any enhancement of the mass-loss rate shortly before explosion.

The distribution of CSM around RSG progenitors is likely highly asymmetric.  Spectropolarimetry of Type II SNe indicates departures from spherical symmetry in the emitting region, though disentangling asymmetry in the CSM from asymmetry in the explosion itself is difficult \citep{wang2008spectropol, vasylyev2023ixf}. Additionally, some SNe show evidence for disk-like or bipolar CSM distributions, possibly arising from binary interaction or rotation-induced asymmetries in the mass-loss process. Understanding the geometry and dynamics of RSG mass loss in the final phases of evolution remains an active area of research, with important implications for interpreting progenitor properties and connecting SNe to their massive star progenitors.

The presence of CSM also complicates the interpretation of progenitor masses derived from pre-explosion imaging. If a significant fraction of the progenitor's bolometric luminosity is reprocessed by circumstellar dust, the observed optical and near-infrared luminosity may significantly underestimate the true stellar luminosity. In other words, if the CSM is dense enough to lead to the formation of narrow lines in the SN spectrum and can substantially enhance the early luminosity in the light curve, it is analogous to LBV mass loss levels and is likely to have a lot of dust that can potentially cause extinction.  Future observations with JWST, which can measure the reprocessed infrared emission directly rather than requiring it to be extrapolated, may help resolve these ambiguities and provide more accurate constraints on RSG progenitor properties.

\subsection{The RSG progenitor of SN~2023ixf}

SN 2023ixf was discovered on 2023 May 19 by \citet{itagaki2023transient} in the nearby galaxy M101, at a distance of $6.85 \pm 0.15$\,Mpc -- the closest supernova of the last decade. Its proximity and early discovery triggered extensive multiwavelength follow-up, making it one of the best-characterized core-collapse supernovae to date \citep[see][and refs within]{jacobson2025ixf}. Given the SN location in M101, a well-imaged, nearby spiral galaxy, the progenitor itself was identified in pre-explosion imaging from both \textit{HST} and \textit{Spitzer}, as well as ground-based near-IR ($JHK$) photometry spanning years to decades before explosion \citep{Jencson2023, Ransome2024}. This multi-wavelength SED allowed the progenitor to be quickly confirmed as an RSG. 

\begin{figure*}
    \centering
    \includegraphics[width=\linewidth]{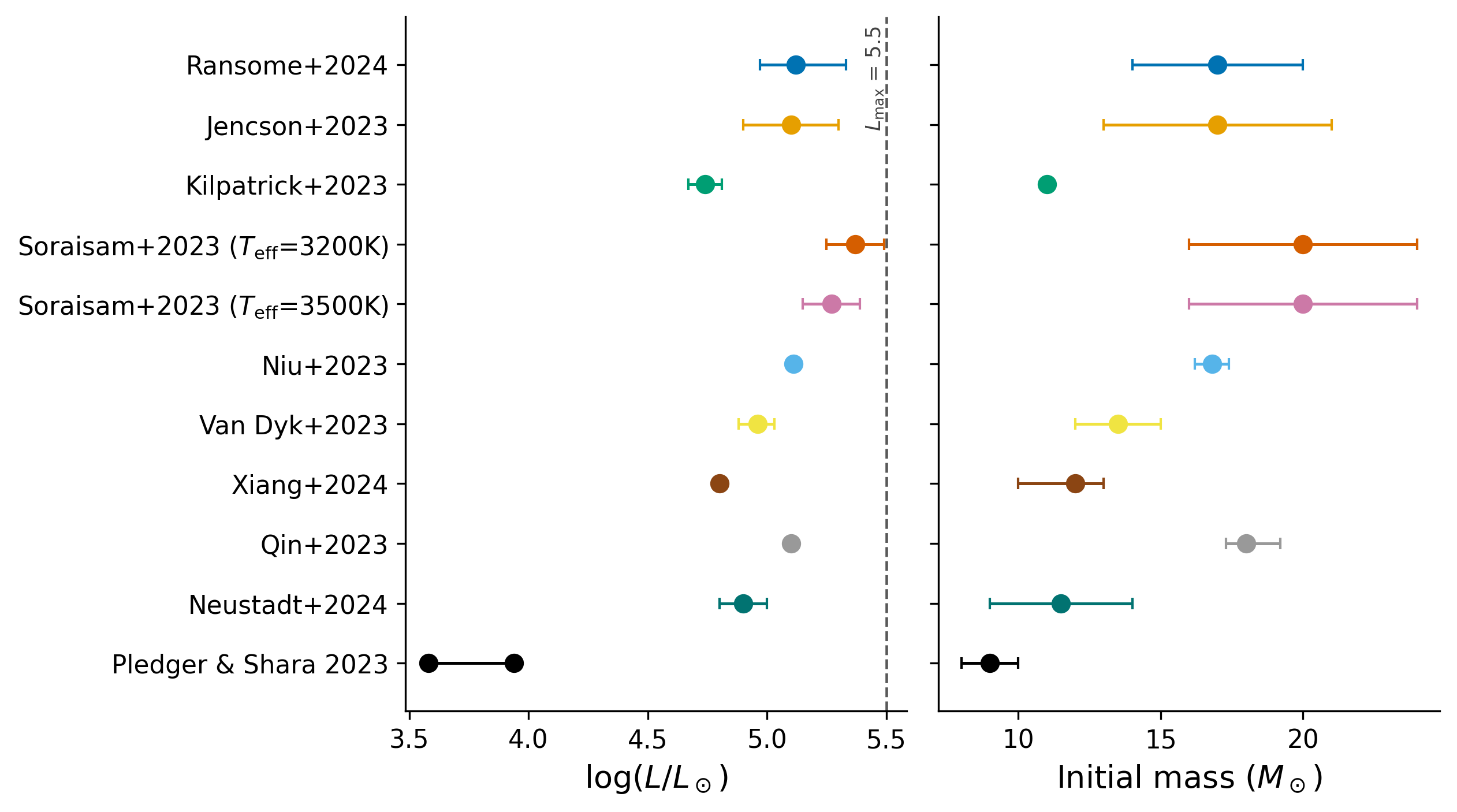}
\caption{Comparison of literature estimates of the SN\,2023ixf progenitor's bolometric luminosity (\textit{left}) and initial mass (\textit{right}). The dashed line in the left panel marks $\logL = 5.5$, the approximate maximum luminosity observed for RSGs locally (the Humphreys--Davidson limit; Section~\ref{sec:hdlimit}). Since \citet{Pledger2023} do not quote a bolometric luminosity for the progenitor candidate, reporting only an extinction-corrected absolute magnitude ($M_{\rm F814W} = -5.11$), we estimate $\logL$ by applying standard bolometric correction assumptions of $BC = 0$ and $BC = 0.9$ to their quoted magnitude. Adapted from Fig. 4 within \citet{Ransome2024}. }
    \label{fig:ixf_lbols}
\end{figure*}

Despite having one of the most extensive datasets of any SN progenitor to date, SED modelling of the photometry by different groups gives luminosities and masses that vary considerably, as shown in Figure~\ref{fig:ixf_lbols}. Reported masses span roughly $11$--$20\,\Msun$, and the corresponding luminosities span almost the entire $L_{\rm bol}$ range observed in the local RSG population. That factor-of-two spread in mass reflects how sensitive derived parameters are to modelling choices (e.g. dust composition, extinction, bolometric corrections) rather than to the underlying photometry, underscoring just how uncertain progenitor properties derived from pre-explosion imaging can be. Most notably, the one luminosity estimate based on \textit{I}-band (F814W) photometry lies well below the rest of the sample \citep{Pledger2023}. This is a particularly instructive case. As discussed in Section~\ref{sec:rsgproblem}, most SN progenitor luminosities in the literature are estimated from this waveband alone, and it may lead to a systematic underestimate of $L_{\rm bol}$.

The progenitor's SED was also extremely dusty. Its near-IR colors were red ($J-K_s \sim 1.6$\,mag), and despite $\sim$13\,years of deep optical monitoring with Pan-STARRS, the star was never detected at $grizy$ wavelengths -- it was visible only in the near- and mid-IR \citep{Jencson2023, Ransome2024}. That level of self-obscuration, together with growing evidence that the surrounding CSM was not spherical, suggests the progenitor's late-time mass loss was both substantial and geometrically complex. Yet the pre-explosion light curves show no sign of an event such as an outburst responsible for the CSM, with the \textit{Spitzer} photometry covering nearly two decades before explosion. The near-IR photometry extends to within 10 days of the SN and shows only the star's normal pulsation cycle, and optical monitoring places comparable limits \citep{Jencson2023, Neustadt2024}. Whatever formed the CSM around 23ixf did not produce a bright eruption.

Spectra of the SN itself showed clear evidence for confined, dense CSM. Narrow line features faded within a few days, and the mass-loss rates inferred from early spectroscopy and light-curve modelling reach $10^{-3}$--$10^{-2}$\,\Myrunit{} \citep{Bostroem2023, jacobson2025ixf}, well above the value derived from the progenitor SED. The CSM mass is nonetheless small, $\lesssim$0.1\,\Msun, and the spectra suggest it is aspherical \citep{smith2023ixf, vasylyev2023ixf}.

Light-curve modelling constrains the envelope independently. \citet{Hsu2025} used the observed pulsation period as an additional constraint to break the usual degeneracy between progenitor radius, ejecta mass and explosion energy, and found that the progenitor requires a large radius ($\gtrsim$950\,\Rsun) and a low mass hydrogen envelope ($\lesssim$3\,\Msun) despite a high initial mass ($\gtrsim$17\,\Msun). If this is right, of order 10\,\Msun{} was removed before explosion. That is more than the CSM contains and far more than any standard RSG wind can remove (see Section \ref{sec:massloss}). It has been suggested that the envelope may instead have been stripped prior to explosion by another route, such as a binary companion \citep{Hsu2025}.

SN\,2023ixf illustrates an important point about mass loss in RSGs that recurs throughout this review, namely that the CSM detected around a star at explosion and the total mass it lost over its lifetime are not necessarily produced by the same mechanism. For 2023ixf, the CSM is confined and asymmetric, and it is enough to shape the SN's early appearance (e.g. narrow spectral lines), but it contains too little mass, and was ejected too recently, to have altered the star's evolutionary path. The envelope deficit implied by \citet{Hsu2025}, by contrast, is an order of magnitude larger and must have been removed well before the final years probed by the CSM and progenitor SED, pointing to a different mechanism entirely, such as binary stripping. Even for the best-observed RSG progenitor of the modern era, no single diagnostic -- pre-explosion photometry, early spectroscopy, or light-curve modelling -- captures the star's full mass-loss history. Instead, each probes a different epoch and radius, and only together do they show that near-death CSM production and lifetime envelope stripping can be entirely decoupled. 

\section{Discussion and Future Directions}

While there are persistent controversies about the evolution of RSGs to SN, there is also an emerging trend. Measurements indicate that RSG winds are considerably weaker than the prescriptions still embedded in most stellar evolution codes, and that the total mass shed by normal winds during the full RSG phase is small. Single stars in the RSG mass range should therefore retain their hydrogen envelopes and die as Type~II-P supernovae, and the observed distribution of progenitor luminosities is consistent with exactly that once sample size and systematics are accounted for. 

The reduced mass-loss rates of RSGs may introduce another challenge to single-star evolution: with lower $\dot{M}$, some single-star models predict RSGs should be found at luminosities above the observed upper limit of $\log L/L_\odot \sim 5.5$ \citep{zapartas2025}, since less envelope is stripped before core collapse, though this is not ubiquitous \citep[e.g.][]{renzo2017systematic, Beasor2021}. This upper limit is often identified with the HD limit, but the HD limit itself cannot be explained by RSG wind mass loss --- stars above it never evolve to the red supergiant stage in the first place, so its metallicity invariance requires some other mechanism, traditionally LBV-type instability/eruptive mass loss, which is plausibly insensitive to metallicity \citep{smith2006eruptive, humphreysdavidson2025}. The observed fraction of CCSNe that are stripped-envelope supernovae cannot be produced by RSG winds at their measured strength, which points to binary interaction rather than to winds. And the CSM inferred from narrow lines in SN spectroscopy and early light curves demonstrates that something does remove material in the final years before collapse, on a timescale too short to affect the evolution but with enough mass to change what the supernova looks like. Weak steady winds, episodic or eruptive events near the end, and binarity throughout for a significant fraction of RSGs are what the observations currently support.

Several specific observations would test this picture. A confirmed Type~II-P progenitor with a well-determined initial mass above 20\,\Msun{} would potentially resolve the discrepancy. Direct detection of an RSG eruption in a long-baseline photometric survey would settle the question of whether such events occur at the rates needed to alter the evolutionary path of an RSG. Conversely, a null result from LSST monitoring over a full survey baseline would place a meaningful upper limit on eruption frequency and force the remaining mass loss onto binary channels. 

JWST can measure full progenitor SEDs rather than extrapolating from one or two optical bands, and can revisit the failed-supernova candidates, ALMA and VLTI resolve wind structure and the acceleration zone in nearby objects, Gaia continues to improve RSG distances and constrain binarity, LSST provides the long-baseline variability statistics that the eruption question requires, and the ELT will extend spectroscopy of individual RSGs well beyond the Local Group, allowing the metallicity dependence of mass loss to be tested over a far wider range than the factor of five currently accessible. On the theoretical side, the most valuable development would be a wind-driving model complete enough to be implemented in an evolution code, so that mass loss is calculated rather than prescribed.

\vspace{6pt}

\funding{E.R.B. is supported by a Royal Society Dorothy Hodgkin Fellowship (grant no. DHF-R1-241114).}

\acknowledgments{I would like to thank Nathan Smith, Ben Davies and Wasif Shaqil for providing helpful comments on the manuscript. I would also like to thank the three anonymous referees whose comments improved the article. Finally, I would like to thank Jacco van Loon for inviting me to write this review article. }

\conflictsofinterest{The author declares no conflicts of interest.}

\appendixtitles{no}

\begin{adjustwidth}{-\extralength}{0cm}

\reftitle{References}

\bibliography{references}

\PublishersNote{}
\end{adjustwidth}
\end{document}